\documentclass[twocolumn, showpacs,preprintnumbers,amsmath,amssymb]{revtex4}
\usepackage{amsmath}
\usepackage{amssymb}
\usepackage{graphicx}
\input epsf
\usepackage{dcolumn}
\usepackage[pdftex,
            pdfauthor={A.V. Yurov, V.A. Yurov},
            pdftitle={The Day the Universes Interacted: Quantum Cosmology without a Wave function},
            pdfsubject={},
            pdfkeywords={Quantum Mechanics, Quantum Cosmology, Classical Cosmology, Phantom fields, Many Interactive Classical Worlds},
            pdfproducer={Latex with hyperref},
            pdfcreator={pdflatex}]{hyperref}
\hypersetup{colorlinks=true, citecolor=blue, linkcolor=red}
\newcommand{\beq}{\begin{equation}}
\newcommand{\beqn}{\begin{equation*}}
\newcommand{\enq}{\end{equation}}
\newcommand{\enqn}{\end{equation*}}
\newcommand{\eb}{{\rm e}}

\newcommand{\R}{{\mathbb R}}

\newcommand{\Z}{{\mathbb Z}}

\renewcommand{\Im}{\text{\rm Im}}

\newcommand{\h}{\hbar}
\newcommand{\p}{{\mathbf p}}

\begin{document}
\allowdisplaybreaks
\title{The Day the Universes Interacted: Quantum Cosmology without a Wave function}
\author{A.V. Yurov}%
\email{AIUrov@kantiana.ru}
\affiliation{Immanuel Kant Baltic Federal University, Institute of Physics, Mathematics and Informational Technology,  Al.Nevsky St. 14, Kaliningrad, 236041, Russia}
\author{V.A. Yurov}%
\email{vayt37@gmail.com}
\affiliation{Immanuel Kant Baltic Federal University, Functionalized Magnetic Materials for Biomedicine and Nanotechnology Center, Institute of Physics, Mathematics and Informational Technology, Al.Nevsky St. 14, Kaliningrad, 236041, Russia}

\date{\today}

\begin{abstract}
In this article we present a new outlook on the cosmology, based on the quantum model proposed by M. Hall, D.-A. Deckert and H. Wiseman \cite{HDW}. In continuation of the idea of that model we consider finitely many classical homogeneous and isotropic universes whose evolutions are determined by the standard Einstein-Friedman equations but that also interact with each other quantum-mechanically via the mechanism proposed in \cite{HDW}. The crux of the idea lies in the fact that unlike every other interpretation of the quantum mechanics, the Hall, Deckert and Wiseman model requires no decoherence mechanism and thus allows the quantum mechanical effects to manifest themselves not just on micro-scale, but on a cosmological scale as well. We further demonstrate that the addition of this new quantum-mechanical interaction lead to a number of interesting cosmological predictions, and might even provide natural physical explanations for the phenomena of ``dark matter'' and ``phantom fields''.
\end{abstract}

\maketitle
\section{Introduction} \label{sec:Intro}

Does the quantum mechanics play any significant role in the contemporary cosmological dynamics? Could our universe as a whole behave as a quantum object? On a first glance, these questions make little sense: after all, the observable universe is an absolutely gigantic object, with the radius of the horizon estimated to be around $10^{28}$ cm. Instead, we are accustomed to encountering the quantum-mechanical phenomena on the atomic scales and below, and almost never -- on the macroscopic scales and above. Of course, there are exceptions: the particular large-scale structure of the universe bears an indelible insignia of ancient quantum fluctuations. Still, even these fluctuations are only noticeable because they have been disproportionately blown up in the course of early cosmological inflation. In other words, the inflation in the young universe has served as a colossal amplifier of the quantum phenomena from the microscopic straight up to cosmological scales. But in the absence of anything comparable to the inflation \footnote{And significantly far from the initial singularity, where the quantum phenomena has to be taken into account.} the universe at large appears to be essentially indifferent to the laws of quantum mechanics. Or is it?

One of the people who invented the idea that the universe as a whole can (and probably should) be treated as a quantum object was Pedro F. Gonz\'alez-D\'iaz \cite{Pedro}. One of the authors (AYu) had a chance to work together with Pedro, and can attest to Pedro's amazing capacity for generating very bold and novel ideas. AYu remembers one particular conversation, when PGD had first proposed the idea that the quantum effects might play a role on the cosmological scale. Initial reaction of AYu was admittedly rather sceptical: why would the quantum effects, negligible already on the scales of small bacteria ($\approx 1 ~\mu$m) might suddenly show up on the scales of about 10 Gly ($10^9$ light years)? PGD came up with a following answer: yes, the quantum effects are all but absent on the macroscopic scale. But this is simply a result of the decoherence phenomena, which suppresses the non-diagonal elements of the density matrix of a quantum system, and which manifests itself whenever that quantum system has any sort of interaction with the environment. But this mechanism does not work for the universe itself, for the universe has {\em no} outside environment, has nothing to interact with and is therefore immune to the effect of decoherence. Therefore, argued PGD, the universe cannot be treated as a classical object, because it can never cease to be a quantum one!

In this article we are going to develop the idea of quantum effects influencing the long-term cosmological dynamics, but we will go by a slightly different route. The key idea here would be not a decoherence (or a lack thereof), but instead a recent article by Hall, Deckert and Wiseman \cite{HDW}, which proposed a new and very novel way to formulate the quantum mechanics. After a little consideration, we have noticed that the method, proposed by Hall, Deckert and Wiseman allows for a very natural extension to the Friedmann cosmology, provided one indeed treats the universe as a quantum object, in full accord with the idea of PGD.

Once we started exploring this idea we have immediately noticed something very unexpected. It is always important to keep in mind that the cosmology, for all its seeming speculativeness, is still a branch of physics. Therefore, all of its ideas, no matter how crazy looking, have to permit for experimental  (i.e. observational) verification. What we have noticed when we started working with our new idea is the following remarkable fact: in our model the quantum effects were inducing the dynamics identical to those produced by the hypothetical phantom fields. To be more precise, during a certain phase of its evolution the universe experiences the phantom acceleration (i.e. $\ddot H >0$ whereas the parameter of state $w<-1$), which proves to be an entirely quantum phenomena \footnote{Fascinatingly, PGD who was written a number of articles about the phantom cosmology, was of the opinion that that the phantoms, if they do exist, has to be treated as a quantum, not classical phenomena -- see, for example, \cite{Pedro_Siguenza}, \cite{Yurov_Pedro}, \cite{Pedro_Perez}.}. The phantoms appeared to be a very unexpected but by no means unwelcome consequence of our proposed theory, so it makes sense to stop and discuss them in a little more detail.

If we are to create a list of the most popular topics in the contemporary cosmology, the term ``phantom fields'' would surely wind up somewhere on the top, being one of the most disputed and discussed subjects today. Taking, for example, all the articles published in ArXiv under the {\em astro-ph} rubric, and choosing among them the ones that contain the word ``phantom'' in the title, one will come out with more than 260 hits (and expanding this search to the {\em hep-th} rubric nearly doubles the number). If we then set our mind to studying these articles, we will soon realize that the majority of them follow a very specific guideline: taking a familiar cosmological model, adding to it a ``phantom field'', solving the resulting equations (analytically if possible, numerically if not) and ruminating over the result. What is usually missing is a discussion of the physical nature of the proposed fields, which, taking into account the singularly mind-boggling properties these fields are supposed to possess (i.e. the wanton violation of the weak energy condition, the negative value of the kinetic term for the scalar field, etc), makes the entire enterprise seem rather  dubious from a physical point of view. Of course, the usual retort to this objection would be a nod to the observational data as an undubitable proof of existence of a hidden type of matter suitably dubbed ``dark energy'' \cite{Perlmutter},\cite{Riess}. We know it exists, we know it constitutes more than 68\% of all matter in the observable universe and, finally, we also know that its parameter of state $\omega$ {\em might} lie just below the vacuum energy threshold $\omega=-1$ \cite{Ade}. Hence, there is a solid chance that at least some of the ``dark energy'' is attributable to the ``phantom fields'' (i.e. fields with $\omega < -1$), so it is only natural to try out the various models that incorporate such fields. Sure, their real {\em cause} remains a mystery, but shouldn't we at least try to guess their {\em effect}?  While this mode of reasoning indeed has some optimistic charm, it unfortunately runs dangerously afoul of the normal scientific approach. We don't mean to imply there is something inherently wicked and wrong with occasional forays into various possible models, done just to see what new effects and phenomena one might uncover; after all, the authors themselves had dabbled in their fair share of unusual cosmologies (such as the universes with the varying speed of light \cite{YY09} and brane worlds \cite{YY05}), and even did some treatises on various aspects of the phantom cosmologies (from the universe-encompassing phenomena of the ``Big Trip'' \cite{AYY12} to the problem of proliferation of Boltzmann brains within the universe filled with the phantom fields \cite{AYY16}). But to have almost the entire field of research succumb to this ``throw it at the wall and see what sticks'' approach is truly utterly frustrating. Because, to put it bluntly, the science simply does not work this way. In fact, if one is is to put an analogy with a cosmological breakthrough of a similar calibre, a Penzias and Wilson's discovery of the cosmic microwave radiation (CMR) \cite{Penzias_Wilson} comes to mind. But in order to make the analogy complete imagine this discovery accompanied not by the corresponding article by Dicke et al. \cite{DPRW}, or any other work providing the explanation of the phenomena by drawing the relationship to the already familiar Big Bang theory. No, instead picture the cosmologists viewing the apparently ubiquitous background radiation as a byproduct of equally ubiquitous and ever-present ``luminous'' matter, and then dedicating article after article to an ambitious tusk of construction of an apt model of the universe filled with such a thing. Imagine, how much time and energy could have been wasted in such a pursuit... Similarly, the understanding we now have of the later stages of stellar evolution in many ways stems from the fact that the research team that has first detected the pulsar signature \cite{HBPSC} chose to look for its explanation in the predictions of the fundamental theoretical physics (i.e. in the behaviour of strong magnetic fields of rapidly rotating neutron stars), and not in the realm of the hypothetical extraterrestrial interstellar communications (although  the team actually gave the signal a codename ``LGM-1'', which of course stands for ``little green men'').

To put it in other words, what we have now is a proliferation of the articles that attempt to explain what the phantom fields {\em do}, a drought of those that explain what these fields {\em are} -- and practically none that have a benefit of showing where these fields would {\em fit in} in the known laws of nature.\footnote{The same can also be said about the ever-growing in both popularity and volume subject of the modified gravity.} So, we have to ask ourselves: is there any way out of this self-imposed predicament? If the history of science (e.g., the aforementioned discovery of CMR) is any indication, the solution to the conundrum of the ``phantom fields'' should be looked for at the most fundamental laws of nature. How deep shall we look? Apparently, the judge of that would be the very properties of these fields. Since they are so radically different from literally every other type of matter (be it dust, radiation or even a vacuum energy), it is a safe assumption that to understand their physical nature we should start at the vary basic level -- where a {\em quantum mechanical} laws reign supreme. And it may be that we already have all the necessary ingredients for that.

Enter the 2014's article \cite{HDW} by M. Hall, D.-A. Deckert and H. Wiseman. In this article the authors have presented a very novel outlook on the problem of a deterministic interpretation of the quantum mechanics. Two such interpretations were known. First of all, there was the de Broglie-Bohm wave-pilot interpretation (dBB) \cite{Bohm}, which describes the behaviour of the quantum particles as the one that obeys the classical Newtonian mechanics with a ``twist'': the equation of motion includes a contribution of an additional field  -- a ``quantum potential'', which depends on the absolute value of the Schr\"odinger's wave function, and is wholly responsible for all the marvelous quantum-mechanical idiosyncracies. Secondly, there was the famous many-worlds interpretation of Hugh Everett III \cite{Everett}, postulating that all the outcomes (no matter how improbable) of any given quantum-mechanical observation are actually physically realized -- each in its own separate ``universe'' peacefully coexisting in a broader multiverse; the very act of observation simply slips us into one of those universes. Both of those interpretations have their own benefits; but both although have their own problems. For example, dBB has to postulate the existence of a certain field that directs and controls the movement of a given quantum particle, but is not directly produced by it -- or any other particle for that matter. The many-worlds interpretation in its turn had to work with and predict the existence of {\em uncountably} many ``quantum worlds'' (to fill up the entire Hilbert space), including very bizarre ones. Furthermore, the worlds themselves do not exert any influence on each other but apparently exist for the sole purpose of accepting the stray visitors after they conducted their ``observations'' -- like a chain of free motels at a sports event...

Now, the crux of the matter is that both of these interpretation has to produce one the same quantum mechanics we know. Therefore, there should theoretically exist a way to combine these two seemingly incompatible world-views. In essence, that is exactly what the authors of \cite{HDW} did. They started out with the many-worlds prediction of an existence of multiple classical worlds. First thing they did was to radically cut down the number of these worlds to just $N \in \Z$ (so there should be at most countably many of them). Furthermore, these classical worlds should not just co-exist with each other, but should in fact {\em interact}, the mechanism of interaction being the de Broglie-Bohm mechanism; in fact, the limit $N \to \infty$ should produce {\em exactly} the same potential as in the dBB framework, except this time it will not be a by-product of some arcane ``external'' field, but a result of the repulsive forces between the neighbouring classical worlds. So, we end up with a picture that successfully combines the two seemingly incompatible interpretations and at the same time actively avoids the weaknesses of both of its predecessors -- it literally takes the best of both worlds -- or, in this case, ``the best of both multiverses''.

We will talk more about this interpretation (known as the interpretation of many interacting worlds a.k.a. MIW) and how it can derived using the dBB formalism in Sec. \ref{sec:MIW}, but right now we are interested in just two of its most important predictions. First of all, MIW predicts that the force of interaction between the different ``worlds'' is repulsive and that it decreases monotonously with the relative distance between them. This implies that the reason the quantum phenomena are so fragile has nothing to do with a ``collapse of a wave function'' (whatever that means) -- in fact, such an object as a wave function is inessential and can be completely avoided in the MIW formalism. No, the existence of quantum phenomena relies solely on the mutual positions of the neighbouring ``worlds'' -- when they are sufficiently close, the quantum potential is alive and kicking; when they depart, the quantum potential abates and the particles become effectively classical again. There is no such thing as ``decoherence''. On the other hand, the similar arguments can be applied to not just one observable particle (and its corresponding ``world'', aptly called in \cite{HDW} the ``world-particle''), but also to the progressively larger groups and structures made of said particles; if the ideology of MIW is correct and there is no decoherence, such macroscopic objects should also be responsive to the quantum-mechanical effects, produced by their quantum ``shadow copies'', provided, that is, that the positions of the particles in the ``doppelg\"anger worlds'' are sufficiently close to those in the observable system.

Naturally, for an ordinary macroscopic object the force exerted by this type of interacting is expected to be vary small. Indeed, if we estimate the probability that the majority of atoms in a human body gets sufficiently close to its its ``doppelg\"angers'', it would be inversely proportional to the whole number of such atoms and so be roughly equal to $10^{-27}$. This, of course, explains why the quantum mechanics and its phenomena are mostly relegated to the domain of very small (atomic-sized) objects. However, unlike the {\em macroscopic} objects (e.g., protists, potatoes, people, planets etc.), on the cosmological scale of entire Friedmann universe, the impediment disappears, owing to the fact that the cosmological dynamics of a Friedmann universe is completely equivalent (with one minor assumption regarding the spectrum of possible energies -- see Sec. \ref{sec:MIU}) to a dynamics of a {\em single} material point situated on a surface of a homogeneous and isotropic  gravitating sphere.


Here we arrive at a second important consequence of MIW: the quantum objects strictly obey the Newtonian physics, just with an additional force induced by the world-to-world interactions. In fact, from the point of view of the Hamiltonian formalism, the quantum physics merely requires an addition of one special term to the Hamiltonian. But it is important to remember that the Einstein-Friedman equations that govern the evolution of the homogeneous isotropic universe can also be {\em derived from the Newton equations} describing the adiabatic expansion of a homogeneous isotropic 3-dimensional sphere of matter -- the only contribution required by the General Relativity and absent from the Newtonian physics being the limitation on the possible spectrum of the curvature constant $\kappa$ (in fact, $\kappa={-1,0,1}$, which corresponds to the open, flat and closed universes). It is also a rather straightforward job to write down a corresponding Hamiltonian (which, of course, corresponds to the one, produced via the General Relativity formalism). If our understanding of the quantum mechanics via MIW is indeed correct, then in addition to the observable universe there should exist a multitude (perhaps even countably many) of shadow universes, that can interact with ours via a quantum potential when these shadow universe are ``close enough'' -- but this time instead of the relative positions of the universes (and yes, this time we are indeed talking not about the ``worlds'', but ``universes'', and one can not really define a ``position of a universe'') -- we have to compare their ``states'', which geometrically can only mean measuring how similar (or dissimilar) their {\em sizes} are, i.e. we have to measure their relative {\em scale factors}. After all, it is the scale factor that replaces the {\em single} coordinate in the Einstein-Friedman equations written in the Hamiltonian formalism, so it makes sense to use it in the MIW extension for cosmology. However, one has to be very careful when implementing this idea. While a construction of the corresponding equations (that effectively combine the Einstein-Fridman equations with the quantum interaction) is straightforward enough (as will be demonstrated in Sec. \ref{sec:MIU}), it is some of the properties of their special solutions that that have to be pondered. For the sake of simplicity, let us consider a special case of just two interacting universes. If we don't impose any sort of initial conditions on their dynamics, it is not impossible to obtain a special solution for the scale factors of these two universes with one of them turning to zero at some moment of time $t_0$ while the other one stays strictly positive both at and in an open neighbourhood of $t_0$ -- in fact, in Sec. \ref{sec:MIU} we show that such solutions indeed exist (for the special case of two empty universes, filled solely with the quantum potential). Now this presents a {\bf problem}. The scale factor turning to zero means that at $t=t_0$ the first universe collapses into a singularity. Hence, at a close vicinity of $t_0$ there should be a brief time period when the entire collapsing (for $t<t_0$) universe is microscopic in size. Such a universe should be under a tremendous influence of the quantum effects. However, we remind the reader that we are working within the framework of MIW paradigm, which explicitly defines all quantum phenomena as manifestations of the repulsive interaction between the neighboring universes -- in our case, the interaction between the first and the second universes. The {\bf problem} then is that this repulsion can only be significant when the relative scale factors are {\em close to each other}, that is, {\em both} of them has to be equal to zero at $t=t_0$ -- and this is definitely not our case.

The only sensible way to eliminate this problem would be to assume that {\em all} scale factors of all available interacting universes should hit zero at exactly the same time, that is -- they all have to be proportional to each other. In other words, all these scale factors shall be proportional to one, special scale factor which we have called the {\em master-factor}. This simple deduction (dubbed the ``master-factor method'') not only resolves the singularity problem we just discussed, but also greatly simplifies the resulting equations, and, most surprising of all, provides a very new outlook on the multiverse itself, asserting that the properties of the observable universe in a strange way depend on a point of view of the observer. If, for example, one was to look at the multiverse from a bird's eye perspective (assuming, for the sake of the argument, that a bird can actually hover over and look upon the entire multiverse), one was to see but one universe, its size determined by the ``master factor'' and its evolution determined by just one ordinary differential equation. However, this is not what the individual observers inhabiting their respective universes will see. These ``down-to-earth'' (or, actually, ``down-to-a-universe'') observers, would have their own view (which can be called a ``frog's'' perspective to distinguish it from the ``bird's'' perspective described above \cite{Tegmark}), with the effective observable scale factor, density, pressure and curvature being potentially quite distinct from those observed by both the ``bird'' and the other ``frogs'' (belonging to the different universes), proving that in some way not only beauty, but the universe itself lies in the eyes of the beholder.

Thus, equipped with both the new modified Einstein-Friedman equations and the master-factor method, in Sec. \ref{sec:MM} we embark on exploration of their possible consequences for the fate of the observable universe. In particular, we demonstrate that even for the simplest case of just two interacting universes, filled with ordinary matter ($\omega=0$), radiation ($\omega=1/3$) and the vacuum energy ($\omega=-1$), by adding the quantum interaction it becomes possible to replicate the effects akin to those of a dark matter and the phantom fields ($\omega < -1$); it might even act to prevent a complete collapse of the universe to a final singularity (Sec. \ref{sec:AdS}), thus eliminating the very point where our knowledge of physics seems to officially break down.

But before we move on to these prospects, it would be a good idea to take a look a the theories that gave rise to them, namely: the De Broglie-Bohm and the Many Interacting Worlds interpretations.

\section{From the de Broglie-Bohm interpretation to Many Interacting Worlds} \label{sec:MIW}

We shall begin by reminding the reader of the basic idea behind the de Broglie-Bohm formalism, since it will serve as a basic framework for the MIW interpretation (for more in-depth analysis of the Bohmian approach see \cite{Durr}). For this end, consider a $D$-dimensional system of $J$ scalar non-relativistic (i.e. possessing the velocities that are much smaller then the speed of light) particles with a total of $n = J D$ degrees of freedom and $\vec q = \{q_i\}_{i=1}^n$ position coordinates. Let $\psi(\vec q, t) = \psi(q_1,...q_n, t)$ be the {\em world wave function} of this system. Apparently, it should satisfy the standard Schr\"odinger equation
\beq \label{schroedinger}
i \h \frac{\partial \psi}{\partial t} = \left(-\sum\limits_{k=1}^{n} \frac{\h^2}{2 m_k} \frac{\partial^2}{\partial q_k^2}+V(\vec q)\right) \psi,
\enq
where $\{q_{3k-2}, q_{3k-1},q_{3k}\}$ are the coordinates of $k$-th particle, $m_{3k-2}=m_{3k-1}=m_{3k}$ is its mass and $V(\vec q)$ is the potential energy of the system in question.

Now, the main idea of the dBB interpretation is that the wave function $\psi$ is a {\em physical field} that controls the dynamics of the particles via their velocities:
\beq \label{q_is_v}
\frac{d \vec q}{d t} = \vec v (\psi).
\enq
If we make this assumption, the shape of the (unknown) function $\vec v$ can be ascertained by the following three physically sound requirements:

\begin{itemize}
\item $v(\psi)$ must be homogeneous of order 0:
\beqn
v(c \psi) = v(\psi), \qquad c > 0,
\enqn

\item equation \eqref{q_is_v} should be invariant w.r.t time-reversal transformations:
\beqn
\begin{split}
t &\to -t, \\
\psi &\to \psi^*,
\end{split}
\enqn

\item equation \eqref{q_is_v} must be Galilean invariant.
\end{itemize}

The first condition -- the homogeneity of $\vec v$ -- stems from a simple fact that the Scr\"odinger equation \eqref{schroedinger} is invariant with respect to multiplication of $\psi$ on an arbitrary constant. Obviously, such a multiplication shall not have any effect on the velocity \eqref{q_is_v} either. In order to carry this simple observation we shall choose a simplest nontrivial homogeneous relationship, so that for a $k$-th particle:
\beq \label{v_is_psi}
\vec v_k (\psi) = \alpha_k \frac{\nabla_k \psi}{\psi}, \qquad \alpha = \text{const}, \qquad \nabla_k = \partial_{q_k},
\enq
where $\alpha_k$ are some (yet unknown) constants, possibly different for different values of $k$. However, we have to keep in mind that while $\psi$ is complex-valued, the velocities $v_k$ shall be real-valued. To see how to accommodate this requirement lets us take a look at yet another property of the Schr\"odinger equation \eqref{schroedinger}, namely: that it is invariant with respect to the time-reversal $t \to -t$ coupled with the conjugate transformation $\psi \to \psi^*$. In terms of \eqref{q_is_v} this symmetry boils down to the condition
\beqn
\vec v_k(\psi) = - \vec v_k(\psi^*),
\enqn
which immediately leads us to the conclusion that it is the {\em imaginary} part of \eqref{v_is_psi} that shall survive:
\beq \label{v_is_psi_really}
\begin{split}
\vec v_k(\psi) &= \alpha_k ~\Im\left(\frac{\nabla_k \psi}{\psi}\right) \\
&= \frac{\alpha}{2i} \left(\frac{\nabla_k \psi}{\psi} - \frac{\nabla_k \psi^*}{\psi^*}\right) \\
&=\frac{\alpha}{2 i} \nabla_k \left( \ln \frac{\psi}{\psi^*}\right),
\end{split}
\enq
where $\alpha_k \in \R$, and it is the value of these constants that we shall deal with next. Since we have assumed that the particles involved are non-relativistic, the equation \eqref{q_is_v} should obey the Galilean invariance principle. Since for any given $k$-th particle the Galilean invariance of the original Schr\"odinger equation w.r.t. transformation $\vec q_k \to \vec q_k + \vec v_0 t$ (where $\vec v_0$ is some constant vector) is ensured by the so-called boost transformation
\beq \label{boost}
\psi \to \psi ~\exp\left(i\frac{m_k}{\h} \vec v_0 \cdot \vec q_k\right),
\enq
upon the substitution of \eqref{boost} into \eqref{v_is_psi_really} we conclude that a $k$-th particle will have its own value of the constant $\alpha_k = \h/m_k$, and so
\beq \label{v_is_psi_really_really}
\begin{split}
\frac{d\vec q_k}{dt} = \vec v_k(\psi) &= \frac{\h}{m_k} ~\Im\left(\frac{\nabla_k \psi}{\psi}\right) \\
&= \frac{\h}{2 i m_k} \nabla_k \left( \ln \frac{\psi}{\psi^*}\right) \\
&= \frac{\h}{i m_k} \nabla_k \left( \ln \frac{\psi}{|\psi|}\right).
\end{split}
\enq

This expression can be greatly simplified if we introduce two real-valued functions $S$ and $R \ge 0$, associated with $\psi$ as
\beq
\psi = P^{1/2} \eb^{i S/\h},
\enq
because then:
\beq \label{q_is_S}
m_k \frac{d \vec q_k}{dt} = \nabla_k S.
\enq

We are now ready to try and understand the actual physical meaning of the dBB interpretation. In order to do this, we simply have to derive the {\em equation of motion} of the  ``world particle'', which is done by finding the value of $\ddot {\vec q}_k$ and utilizing the Schr\"odinger equation \eqref{schroedinger}. After a simple differentiation we will end up with nothing else but a {\em generalization of the Newton equation}:
\beq
m_k \frac{d^2 \vec q_k}{d t^2} = -\nabla_k V(q) - \nabla_k Q(P),
\enq
where the new $P$-dependent function $Q$:
\beq
Q = - \frac{\h^2}{2m_k} \frac{\Delta_k P^{1/2}}{P^{1/2}},
\enq
plays the role of a {\bf quantum potential} (Q-potential), which is responsible for the quantum mechanical peculiarities. The dynamics is completely deterministic, and the probabilities we are so accustomed to see in the context of the quantum mechanics are now simply a manifestation of our lack of knowledge with regards to the exact initial positions of $J$ particles. It is this ignorance and nothing else that {\em forces} us to associate $P=|\psi|^2$ with the probability.

The main problem of the dBB interpretation is that it postulates a new field $Q$, which affects the behavior of the particle, but is seemingly unaffected by the particle itself. Needless to say, it is this indifference which places it in stark contrast with literally any other known physical field.

However, in 2014, the group of M. J. W. Hall, D.-A. Deckert and H. M. Wiseman has proposed a very novel way to resolve this difficulty. Their idea was to combine the dBB approach with the Everett's many worlds interpretation -- the approach called the {\em many classical interactive worlds interpretation (MIW)} \cite{HDW}.

In order to follow this argument let us assume that instead of one unique world particle travelling in the phase space in a supreme loneliness, there are now $N \gg 1$ of them, located at points $\vec x_k(t)$ and having masses $m_k$. As we are ignorant of the actual positions of these worlds, we have to treat their initial positions as random variables, with the distribution probability $P(0,x_n)$, where
\beqn
P(t,q)= |\psi(t,q)|^2,
\enqn
which satisfies:
\beq
P_0(\vec q) \approx \frac{1}{N} \sum\limits_{n=1}^N \delta \left(\vec q - \vec x_n(0)\right).
\enq

We have to understand the dynamics of these worlds. If we require the dBB approach to be just a manifestation of MIW paradigm at the limit $N \to \infty$, then it is natural to seek the equations of motion in the Newtonian form
\beq
\frac{d^2 (\overrightarrow{m_n q_n})}{d t^2} = -\nabla_{x_n} V - \nabla_{x_n} U_N,
\enq
where $U_N$ plays a role of the {\em interworld interaction parameter}, which in the limit $N\to \infty$ becomes the dBB's Q-potential.

Now we end up with a following Hamiltonian system:
\beq
\begin{split}
(\vec x_n)' &= \nabla_{p_n} \mathcal{H}_N(\vec x, \vec p), \\
(\vec p_n)' &= - \nabla_{x_n} \mathcal{H}_N(\vec x, \vec p),
\end{split}
\enq
where the $'$ is the derivative w.r.t. time, the momenta $\vec p_n$ are defined as
\beq
\vec p_n = m_n \vec x_n '
\enq
and the Hamiltonian
\beq
\mathcal{H}_N = \sum\limits_{n=1}^{N} \frac{|\vec p_n|^2}{2 m_n} + \sum\limits_{n=1}^{N} V(\vec x_n) + U_N(\vec x_1,..,\vec x_N)
\enq

By assumption in the limit $N\to \infty$ the MIW model shall become the dBB one, so for $N$ large enough $\vec p_n \approx \nabla_{x_n} S$. Comparing then the average energy per world $\mathcal{H}_N/N$ with the known expression for the quantum average energy we conclude that
\beq
U_N \approx \sum\limits_{n=1}^N \frac{\h^2}{8 m_n} \frac{\nabla P(x_n)}{P(x_n)}.
\enq

So, the shape of the potential depends on the exact form of functions $P(x_n)$. For the sake of simplicity, let us find what it should look like in the simplest one-dimensional case of $N$ interactive worlds, positioned at coordinates $x_1<x_2<..<x_N$. Replacing the function $P$ with its smooth interpolation $P_a$ (assuming that $N \gg 1$), for a sufficiently slowly varying function $\phi$ we should have:
\beq
\begin{split}
\frac{1}{N} \sum \limits_{n=1}^N \phi(x_n) &\approx \sum \limits_{n=1}^{N-1} \int \limits_{x_n}^{x_{n+1}} dq P_a(x_n) \phi(x_n) \\
&= \sum \limits_{n=1}^{N-1} (x_{n+1} - x_n) P_a(x_n) \phi(x_n),
\end{split}
\enq
which means that for a one-dimensional case of $N$ worlds with slowly-varying distance the ansatz distribution and corresponding interworld potential are:
\beq
\begin{split}
P &= \frac{1}{N(x_{n+1}-x_n)} \\
U_N &= \frac{\h^2}{2m} \sum\limits_{n=1}^N \left(\frac{1}{x_{n+1}-x_n}-\frac{1}{x_n-x_{n-1}}\right)^2,
\end{split}
\enq
where for simplicity we defined $x_0=-\infty$ and $x_{N+1}=+\infty$.

A note before we move on... In a recent paper by I. McKeague, E. Pek\"oz and Y. Swan \cite{MPS} a more general case has been studied, in which the interworld potential has the form
\beq
\begin{split}
\ U_N = \frac{\h^2}{2m} \sum\limits_{n=1}^N &\Big(\frac{1}{B(x_{n+1})-B(x_n)}-\\
&- \frac{1}{B(x_n)-B(x_{n-1})}\Big)^2 (B'(x_n))^2,
\end{split}
\label{Tema0}
\enq
where the function $B$ is a continuously differentiable non-decreasing function with finitely many critical points, such that $B(x_0)=-\infty$ and $B(x_{N+1})=\infty$. However, it can be shown that our qualitative results remain the same for any power law $B(x)=x^n$, $n>0$, so we'll restrict our attention to the case $B(x)=x$.

What is most staggering about the MIW approach (just like the Everett's Many World Interpretation) is the following often overlooked fact: it makes the mysterious concept of a ``wave function's collapse'' completely obsolete. The immediate corollary then would be that the quantum mechanical phenomena is not something solely restrained to the microscopic scale!

As we have seen, in terms of the MIW approach the quantum mechanics appears to be an extension of a Newtonian mechanics, where a particle ends up being surrounded by $N-1$ of its ``doppelg\"angers'', each one contributing its share into the combined ``shadow'' potential, that is ultimately responsible for the quantum mechanical phenomena.
But then the same process can be performed for the macroscopic objects; the size doesn't matter as long as there different copies of the object, interacting with each other. The only thing that matters is how close in the phase space these ``shadow copies'' are.

But even that is not all, because this very idea can also be extended to the classical cosmology and naturally applied to the {\em universe itself}.

\section{Many Interacting Universes} \label{sec:MIU}

Indeed, recall that Friedmann equations serves as a simple generalization of the Newtonian mechanics. First of all, let us consider
a material point of unit mass, located on a surface of a three-dimensional gravitating sphere of radius $a(t)$ (uniformly and isotropically filled with matter of density $\rho(t)$ with pressure $p(t)$),  with the overall classical energy $E=-\kappa c^2/2$ (where $\kappa=-1,0,1$). Assuming the cosmological constant $\Lambda = 0$, for any object at the boundary of the sphere the conservation of energy law becomes the first Friedmann equation \cite{Barrow_Tipler}:
\beq
\frac{\dot a^2}{2} -\frac{4 \pi G}{3} a^2 \rho = - \frac{\kappa c^2}{2},
\enq
whereas the adiabaticity of the expansion of the universe implies that $d W = -p dV$, i.e.
\beq
\dot \rho = -3\left(\rho + \frac{p}{c^2}\right) \frac{\dot a}{a}.
\enq
Combining them together (by differentiating the conservation law) we end up with the second Friedmann equation:
\beq
\ddot a = -\frac{1}{2}\left(\rho +\frac{3p}{c^2}\right) a.
\enq

Therefore, we can account for the quantum interaction in cosmological equations in exactly the same manner as with the Newton equation: introduce a Q-potential, describing the ''interaction'' of N classical one-dimensional material points, interpreting it as an ``interaction'' of N classical 
Friedmann universes. If their states are sufficiently different (they are ``far'' from each other in the phase space), their dynamics should be identical to the one predicted by the classical cosmology. But if their states are close enough, the neighbouring universes should experience a quantum ``repulsion''.

We can derive the Friedmann equations with the Q-potential as follows. We start with the classical Hamiltonian (assuming for simplicity that $8 \pi G/3 =1$ and $c=1$):
\beq
\mathcal{H}(a,\p) = \frac{\mathbf{p}^2}{2} - \frac{1}{2} \rho(a) a^2 = -\frac{\kappa}{2}.
\enq
Using the inverse Legendre transform we can produce the Lagrangian:
\beq
L = \frac{1}{2} \dot a^2 + \frac{1}{2} \rho(a) a^2,
\enq
which, by a courtesy of the Euler-Lagrange equation and the continuity equation on $\dot \rho$, subsequently produces both of the Friedmann equations.

Thus, in order to account for the Q-potential of N interactive Friedmann universes with scale factors $a_n$, densities $\rho_n$ and pressures $p_n$, $n=1,..N$, related via
\beqn
\dot \rho_n = -3(\rho_n + p_n) \frac{\dot a_n}{a_n}.
\enqn
we shall write the new ``Hamiltonian'' $\mathcal{H}_N$ as
\beqn
\begin{split}
\mathcal{H}_N =& \sum\limits_{n=1}^N \frac{\mathbf{p}_n^2}{2} - \frac{1}{2}\sum_{n=1}^{N}\rho_{n}(a_{n})a_{n}^{2} +\\
&+ U (a_1,..,a_{N})+\frac{1}{2}\sum_{n=1}^{N}K_{n} = 0,
\end{split}
\enqn
where $K_n = 0, \pm 1$,
\beq
U(a_{1},..,a_{N}) = \alpha^{2}\hbar^{2}\sum_{n=1}^{N}\left(\frac{1}{a_{n+1}-a_{n}}-\frac{1}{a_{n}-a_{n-1}}\right)^{2},
\label{Tema000}
\enq
and $\alpha^2$ is as yet undetermined constant (as before, we define $a_0=a_{N+1}=\infty$).

Next, we derive the Lagrangian
\beqn
\begin{split}
L &=\frac{1}{2}\sum_{n=1}^{N}\dot{a}_{n}^{2}+\frac{1}{2}\sum_{n=1}^{N}\rho_{n}(a_{n})a_{n}^{2} -\\
&- \alpha^{2}\hbar^{2}\sum_{n=1}^{N}\left(\frac{1}{a_{n+1}-a_{n}}-\frac{1}{a_{n}-a_{n-1}}\right)^{2},
\end{split}
\enqn
...and use the Euler-Lagrange equations
\beqn
\frac{d}{dt}\frac{\partial L}{\partial \dot{a}_{n}}=\frac{\partial L}{\partial a_{n}}, \qquad n = 1,..,N,
\enqn
to obtain the new and improved Friedmann equations:
\beqn
\begin{split}
\ddot{a}_{n} &= -\frac{1}{2}(\rho_{n}+3p_{n})a_{n} -\\
&- \alpha^{2}\hbar^{2}\frac{\partial}{\partial a_{n}}\sum_{k=1}^{N}\left(\frac{1}{a_{k+1}-a_{k}}-\frac{1}{a_{k}-a_{k-1}}\right)^{2}.
\end{split}
\enqn

For a particular example let's take a look at our system for a simplest possible case $N=2$. In it we have just two equations:
\beq
\begin{split}
\ddot{a}_{1} & =-\frac{1}{2}\left(\rho_{1}+3p_{1}\right)a_{1}+\frac{4\alpha^{2}\hbar^{2}}{(a_{1}-a_{2})^3},\\
\ddot{a}_{2} & =-\frac{1}{2}(\rho_{2}+3p_{2})a_{2}-\frac{4\alpha^{2}\hbar^{2}}{(a_{1}-a_{2})^3},
\end{split}
\enq
and the dynamics of the universes are only influenced by the quantum interaction when $a_1 \approx a_2$; outside of this interval, the evolution of either universe is determined (with a good accuracy) by the standard Friedmann equations.

{\bf However.} Left in this form, the Friedmann system produces results of a rather dubious if not unphysical nature. If we look at the general solutions of this system we will immediately notice that the zeroes of the scale factors in two universes do not coincide with other (say, $a_1(t_0)=0$, but $a_2(t_0) \neq 0)$. And this constitutes a problem. When a scale factor approaches zero, that universe must invariably experience a surge of quantum influence. But in our toy model the only source of the Q-potential in one universe is the {\em other} universe. And by our logic, this implies that the scale factors of {\em both} universes has to be very close to each other, i.e. that $a_2(t_0) \approx a_1(t_0) = 0$. So, by necessity, if one universe collapses to a point, the second should too.

Let's see how this problem unfolds for a particular example of two empty universes (with $\rho_1=\rho_2=0$). Let's denote $a_1=x$ and $a_2 = y$. Then the system
\beq
\begin{split}
\ddot{x} &=\frac{4\alpha^{2}\hbar^{2}}{(x-y)^3},\\
\ddot{y} &=-\frac{4\alpha^{2}\hbar^{2}}{(x-y)^3},
\end{split}
\enq
can be integrated:
\beq
\begin{split}
x &= u t + N_{0}+\frac{1}{2\epsilon}\sqrt{4\alpha^{2}\hbar^{2}+2\epsilon^{4}(t-t_{Q})^2},\\
y &= u t + N_{0}-\frac{1}{2\epsilon}\sqrt{4\alpha^{2}\hbar^{2}+2\epsilon^{4}(t-t_{Q})^2},
\end{split}
\enq
where $u, N_0, \epsilon, t_Q$ are constants that satisfy the condition
\beq
\kappa_1+\kappa_2+\epsilon^2+u^2=0.
\enq

\begin{figure}
\begin{center}
\includegraphics[width=0.9\columnwidth]{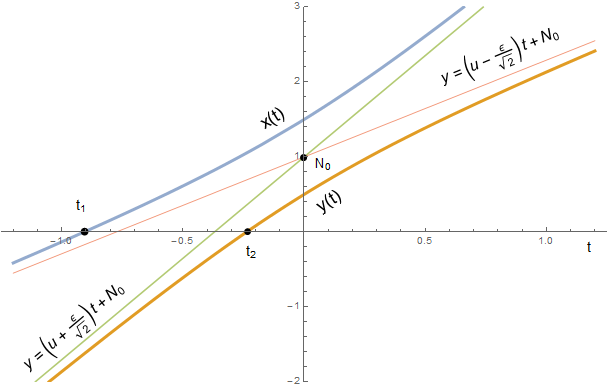}
\caption{\label{fig1} Evolution of the scale factors of two interacting universes. Case $u>\sqrt{\epsilon}/2$.}
\end{center}
\end{figure}

\begin{figure}
\begin{center}
\includegraphics[width=0.9\columnwidth]{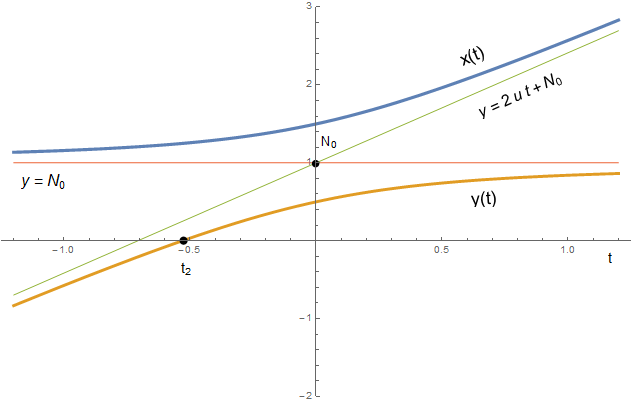}
\caption{\label{fig2} Evolution of the scale factors of two interacting universes. Case $u = \sqrt{\epsilon}/2$.}
\end{center}
\end{figure}

\begin{figure}
\begin{center}
\includegraphics[width=0.8\columnwidth]{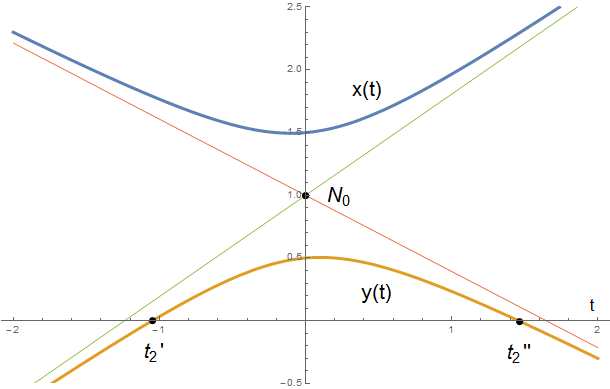}
\caption{\label{fig3} Evolution of the scale factors of two interacting universes. Case $u < \sqrt{\epsilon}/2$.}
\end{center}
\end{figure}

Depending on the relationship between $u$ and $\epsilon$ we get three possibilities for the possible dynamics of $x(t)$ and $y(t)$ (see Figs. \ref{fig1}-\ref{fig3}).

The previous example demonstrated that we have to modify our equations so that {\em all} universes reach singularity at exactly the same time $t_s$. The simplest way to achieve this is by assumption that all scale factors are {\em proportional} to each other:
\beq
a_n(t) = \mu_n a(t), \qquad \mu_n >0, \qquad n=1,..,N,
\enq
where $a(t)$ is the universal scale factor that governs the Q-potential-induced evolution of all universes. For simplicity and also to emphasise its role we will henceforth refer to $a(t)$ as the {\em master-factor}.

Armed with the master-factor, we can rewrite our Q-potential as
\beq
\begin{split}
U & = \frac{\sigma_{N}^2}{a^{2}},\\
\sigma_{N}^{2} &=c^{2}L_{PL}^{2}Z_{N}^{2},\\
Z_{N}^{2} &=\sum_{n=1}^{N}\left(\frac{1}{\mu_{n+1}-\mu_{n}}-\frac{1}{\mu_{n}-\mu_{n-1}}\right)^{2},
\end{split}
\label{Tema00}
\enq
where $L_{pl} = \sqrt{G \h /c^3}$ is the Planck length (it replaces the constant $\alpha$) and $\mu_n$ is the $n$-th universe's constant multiplier to the master-factor.

The corresponding Hamiltonian is then
\beq \label{H}
\mathcal{H}=\frac{1}{2}M_{N}^{2}\dot{a}^{2}-\frac{4\pi G}{3}\sum_{n=1}^{N}\mu_{n}^{2}\rho(\mu_{n}a)a^{2}+\frac{\sigma_{N}^{2}}{a^{2}}+\frac{c^2K_{N}}{2}=0,
\enq
where
\beq
M_{N}^{2}\equiv\sum_{n=1}^{N}\mu_{n}^{2},
\enq
So the entire dynamics of the quantum multiverse is determined by one (sic!) equation \eqref{H}. This overall dynamics is what one would see from a overall, ``birds-eye'' perspective.

On the other hand, the observer who resides in one of those universes will possess what we might call a ``frog's'' perspective (using the term first proposed by Max Tegmark in \cite{Tegmark}), where the dynamics of the ``frog's'' universe is governed by the {\em effective} Friedmann equations
\beq
\begin{split}
\dot{a}_{n}^2 &= \frac{8\pi G}{3} \tilde{\rho}_{n} a_{n}^2 - \frac{\kappa_n c^2}{2}, \\
\dot{\tilde{\rho}}_{n} &= -3\left(\tilde{\rho}_n + \frac{\tilde{p}_n}{c^2}\right) \frac{\dot a_n}{a_n},
\end{split}
\enq
where $\tilde{\rho}_n$, $\tilde{p}_n$ denote the density and the pressure as measured by the ``frog''-ish observer from the $n$-th universe.

For example, consider a ``squeaky clean'' multiverse in which $\rho(\mu_n a)=0$ for all $n$. Then, provided that $K_N <0$ the equation \eqref{H} will have a following nontrivial solution

 \begin{equation}
  a(t)=\frac{c}{M_N}\sqrt{(t-t_{min})^2+\xi_N^2t^2_{PL}},
  \label{Tema1}
  \end{equation}
  where $t_{min}$ corresponds to the time when the master factor reaches its minimal value (and for the further calculations can be set equal to $0$),
  $$
  \xi_N^2=2M_N^2Z_N^2,\qquad t_{PL}=\frac{L_{PL}}{c},
  $$
  and for simplicity we have also made the assumption:
  \beq
  K_N=\sum_{n=1}^N k_n=-1.
  \label{Tema2}
  \enq

Before we move on, it is interesting to point out that the condition \eqref{Tema2} can lead to some curious conclusions. In particular, one can ask the following question: if the number $N$ is known, how many possible configurations of the multiverse would all \eqref{Tema2} to hold? In other words, how many flat ($k_n=0$), open ($k_n=-1$) and closed ($k_n=+1$) can we have so that \eqref{Tema2} remains valid?.. If we restrict ourselves to those numbers $N$ that are odd: $N=2M+1$, $M \in \Z$, then we can estimate the amount of all possible variations as:
 \beq
  \begin{split}
  N(k=0)&=\sum_{m=0}^{M-1}(m+1)C_{2M+1}^{2m+2}C_{2m+2}^{m+1},\\
  N(k=-1)&=\sum_{m=0}^{M}(m+1)C_{2M+1}^{2m+1}C_{2m+1}^{m},\\
  N(k=+1)&=\sum_{m=1}^{M}m  C_{2M+1}^{2m+1}C_{2m+1}^{m}.
 \end{split}
    \label{Tema3}
    \enq
For example, if $M=10$ ($N=21$) we have  $N(k=0)=7,643,699,280$, $N(k=-1)=7,924,966,749$, $N(k=+1)=6,857,074,350$. With the growth of $N$, the parts of universes tend to 1/3. For example. for $N = 1001$, the share of $N (k = 0)$ is  33.3 \%, the share of $N (k = -1)$ is 33.4\%, and the share of $ N (k = + 1) $ is 33.3 \%. The number of corresponding variants grows exponentially fast, say, for $ N = 1001 $, the total number $ N (k = 0, \pm 1) =6,125,509,264 \times 10^{478} $.

After this brief digression into combinatorics, lets return from the ensemble of the universes to a particular one and to the point of view of one of its ``frogs''.

First of all, we note that the Hubble parameter remains the same for all universes:
\beq
H_n=H=\frac{t}{t^2+\xi_N^2 t^2_{PL}}.
\label{Tema4}
\enq
The same cannot be said about the observable properties of the matter; from the point of view of the ``frog'' from the $n$-th universe:
\beq
\begin{split}
{\tilde{\rho}}_n=\frac{3}{8\pi G\mu_n^2}\frac{\left(\mu_n^2+k_nM_N^2\right)t^2+k_n M_N^2\xi_N^2 t^2_{PL}}{\left(t^2+\xi_N^2t^2_{PL}\right)^2},\\
w_n=-\frac{1}{3}\left(1+\frac{2\mu_n^2\xi_N^2t^2_{PL}}{\left(\mu_n^2+k_nM_N^2\right)t^2+k_n M_N^2\xi_N^2 t^2_{PL}}\right),
\end{split}
\label{Tema5}
\enq
with ${\tilde p}_n=w_n{\tilde{\rho}}_n$. The expression for $a_n(t)$  has the form of a simple hyperbola with asymptotes
\beq
y_{\pm}=\pm\frac{c\mu_n t}{M_N}.
\label{Tema-as}
\enq
That is, the smallest possible value $a_n(t=0)$ (here we have assumed that the integration constant $t_Q = 0$) would be equal to $a_n(0)=\mu_n\xi_NL_{PL}/M_N$.

Now, let's once again return back to the ``rebouncing'' model (\ref{Tema1}), this time looking at it from the point of view of three ``frogs'': one from a closed, another from a flat and the last one from an open universe.
\newline

{\bf Case 1}: $k_n=+1$. In this case ${\tilde{\rho}}_n>0$ and ${\tilde p}_n<0$ for $-\infty<t<+\infty$. Accordingly, the $n$-th universe is constantly accelerating, both at the expansion stage and at the compression stage:
\beq
\frac{{\ddot a}_n}{a_n}=\frac{\xi_N^2t^2_{PL}}{(t^2+\xi_N^2t^2_{PL})^2},
\label{Tema6}
\enq
and throughout evolution
\beq
{\tilde{\rho}}_n+\frac{3{\tilde p}}{c^2}<0,
\label{Tema7}
\enq
and (\ref{Tema6}), (\ref{Tema7}) hold for all $k_n$. Interestingly, the weak energy condition in this universe is never violated:
\beq
w_n+1=\frac{2}{3}\frac{(\mu_n^2+M_N^2)t^2+\xi_N^2t^2_{PL}(M^2_N-\mu_n^2)}{(\mu_n^2+M_N^2)t^2+\xi_N^2M^2_N t^2_{PL}}>0,
\label{Tema8}
\enq
because $M^2_N-\mu_n^2>0$. For $|t|\to\infty$, $w_n\to -1/3$, which is obvious since the asymptotically $a_n(t)$ is a  linear function, see (\ref{Tema-as}).
\newline

{\bf Case 2}: $k_n=0$. For this type of universe things starts to get more interesting. $\tilde{\rho}_n$ is now an even function with three extrema (see Fig. 6):
\beq
\begin{split}
\tilde{\rho}_n=\frac{3t^2}{8\pi G(t^2+\xi_N^2t^2_{PL})^2},\\
w_n+1=-\frac{1}{3}\left(1+\frac{2\xi_N^2t^2_{PL}}{t^2}\right).
\end{split}
\label{Tema9}
\enq
From the formulas (\ref{Tema9}) two interesting conclusions can be drawn:

\begin{enumerate}
\item For $t\in(-\xi_Nt_{PL},\,\xi_N t_{PL})$, the universe exists in a phantom zone.

\item At the rebound point ($t=0$), there is a $w$-singularity  ($w_n(0)=-\infty$) \cite{Dabrowski}.
\end{enumerate}

Therefore, we end up with the following conclusion: inside of a flat universe, the Q-potential leads to a double crossing of the phantom threshold and to the emergence of a $w$-singularity.
\newline

{\bf Case 3}: $k_n=-1$. This is the most sophisticated case. First of all, we note that the effective density (that is, from the point of view of the ``frog'') is negative everywhere. This, however, should not discourage us, as it leads to no physical problems. In fact, the negative sign of the density is but an artifact of the negative curvature of the universe:
$$
\tilde{\rho}_n=\frac{3}{8\pi G}\left(H^2_n-\frac{c^2}{a_n^2}\right)<0.
$$
The pressure is more complicated, as it will depend on which of the five intervals
$$
1<\gamma<2,\,\, \gamma=2,\,\, 2<\gamma<3,\,\,\gamma=3,\,\, \gamma > 3,
$$
the value of $\mu_n=M_N/\sqrt{\gamma}$ happens to be in. Because of this, we will omit the corresponding formulas, but will instead confine ourselves to writing down the general expression for the parameter of the equation of state:
\beq
w_n=-1+\frac{2}{3}\left(1+\frac{\xi_N^2t^2_{PL}}{(\gamma-1)t^2 + \gamma\xi_N^2t^2_{PL}}\right).
\label{Tema10}
\enq
The reader will easily build the corresponding formulas for $\tilde{p}$ in each of the five intervals. We note only the following most important observation: since $\gamma > 1$ ($\mu_n^2<M_N^2=\sum_{k=1}^N\mu_k^2$), it follows from (\ref{Tema10}) that $w_n>-1$ for all $t$.

Thus we arrive at the most important conclusion: in a universe in which the Q-potential dominates, phantoms (from the point of view of a ``frog'') arise only if the universe is flat. In other words, if future observations confirm the presence of a phantom component in dark energy and our model is correct, then phantoms will serve as a direct evidence that the we (the ``frogs'') live in a flat universe.

Concluding this section, we recall that as the "original sample" we used the potential (\ref{Tema000}). We have already mentioned that there exists a more general potential (\ref{Tema0}) and that its use does not lead to new results. Now we can easily explain this. Since the general expression for $B(a_n)$ depends on $a^n$,  it can be shown that, within the framework of the master-factor method, potentials of the form (\ref{Tema0})  will lead to the same expression (\ref{Tema00}), just with a different $\sigma_N^2$. In other words, the picture will not change qualitatively.

\section{How does the matter matter?}\label{sec:MM}

Now let us move one step further and add to our picture the matter fields. As before, our description will depend on which of two points of view we will adopt: the point of view of the ``bird'' (which ``observes'' the multiverse in its entirety) or the one of the ``frog'' (which is confined to one of the universes). We will assume that every universe from the ensemble is filled with three types of matter: baryonic matter, photons and vacuum energy. According to the ``bird'', the density of matter in the $n$-th universe has the form
\beq
\rho_n^{(b)}(a_n)=\frac{C^{(b)}_{n(R)}}{a_n^4}+\frac{C^{(b)}_{n(M)}}{a_n^3}+\rho^{(b)}_{n(\Lambda)},
\label{Tema41}
\enq
where    $C^{(b)}_{n(R)}$,  $C^{(b)}_{n(M)}$  and   $\rho^{(b)}_{n(\Lambda)}$  characterize the number of photons, the number of baryons and the density of the vacuum energy as seen by the ``bird'' (which is indicated by the superscript $(b)$ in parentheses). The scale factor $a_n(t)$  is related to the master factor $a(t)$ by the same formula $a_n(t)=\mu_na(t)$, and the value of $\rho^{(b)}_{n(\Lambda)}$ can be positive, negative or equal to zero. This expression we substitute to (\ref{H}) in order to obtain the equation for the master factor:
\beq
\frac{M_N^2{\dot{a}}^2}{2}-\frac{m_N^2}{a}+\frac{\sigma_N^2-r_N^2}{a^2}-\frac{4\pi G}{3}\rho_N a^2+\frac{c^2K_N}{2}=0,
\label{Tema42}
\enq
with
\beq
\begin{split}
m_N^2 &=\frac{4\pi G}{3}\sum_{n=1}^N\frac{C^{(b)}_{n(M)}}{\mu_n},\qquad
r_N^2 =\frac{4\pi G}{3}\sum_{n=1}^N\frac{C^{(b)}_{n(R)}}{\mu_n^2},\\
\rho_N &=\sum_{n=1}^N \mu_n^2\rho^{(b)}_{n(\Lambda)}.
\end{split}
\label{Tema43}
\enq

Now let us look at the same problem but from a ``frog's'' perspective (which, of course, includes us). By analogy with the ``bird's'' formulas, we will denote the corresponding variables by the superscript $(f)$ in parentheses (the previously used $k_n$ are recorded from the point of view of the ``bird'' and therefore must be equipped with an index; $k^{(b)}_n$. Similarly, $a_n$  which we used before, is also defined from the point of view of the ``bird'' and should be written in the form $a^{(b)}_n$. We did not do this in the previous sections to avoid cluttering the formulas). In particular, the parameter with the curvature should appear in the term with curvature as $k_n^{(f)}=-1,0,+1$.  For this it is necessary to renormalize $a_n(t)$ (i.e $a^{(b)}_n(t)$), introducing a new "froggy" scale factor $R^{(f)}_n$:
\beq
a_n(t)=\frac{\mu_n}{M_N}\sqrt{\left|\sum_{l=1}^N k^{(b)}_l\right|}R^{(f)}_n(t).
\label{Tema44}
\enq
After the calculations one gets
\beq
\begin{split}
\left(\frac{{\dot{R}}^{(f)}_n}{R^{(f)}_n}\right)^2=&\frac{8\pi G}{3}\left(\frac{C^{(f)}_{n(RQ)}}{(R^{(f)}_n)^4}+\frac{C^{(f)}_{n(M)}}{(R^{(f)}_n)^3}+\rho^{(f)}_{n(\Lambda)}\right)-\\
&-\frac{c^2 k^{(f)}_n}{(R^{(f)}_n)^2},
\label{Tema45}
\end{split}
\enq
where
\beq
\begin{split}
C^{(f)}_{n(QR)} &=M_N^2\left|\sum_{l=1}^N k^{(b)}_l\right|^{-2}\left(\sum_{k=1}^N \frac{C^{(b)}_{k(R)}}{\mu_k^2}-\frac{3\sigma_N^2}{4\pi G}\right),\\
C^{(f)}_{n(M)} &=M_N\left|\sum_{l=1}^N k^{(b)}_l\right|^{-3/2}\sum_{k=1}^N\frac{C^{(b)}_{k(M)}}{\mu_k},\\
\rho^{(f)}_{n(\Lambda)} &=\frac{1}{M_N^2}\sum_{k=1}^N\mu_k^2\rho^{(b)}_{k(\Lambda)},\qquad k_n^{(f)}={\text {sgn}}\left(\sum_{l=1}^N k^{(b)}_l\right).
\end{split}
\label{Tema46}
\enq

The expressions obtained herein are extremely interesting. We see that all ``frogs'' will receive identical constants characterizing the density of contributions of different matter fields and the same sign of curvature. Virtually all ``frogs'' believe that they live in absolutely identical universes -- the index $n$ in all the ``froggish'' formulas can be removed. And yet, in reality (that is, from the point of view of a ``bird''), the universes are filled with matter in a very different ways and all have different spatial curvatures. This serves as an excellent demonstration for how significantly the points of view of a ``birds'' and ``frogs'' differ.

This circumstance reaches a paramount importance when we take into account the actual observations. In particular, a ``frog'' determines the number of baryons via the astronomical observations (by calculating the average number of stars, galaxies, etc.), and therefore should receive the coefficient $C^{(b)}_{n(M)}$. On the other hand, the dynamics of the universe observed by the ``frog'' is described by the contribution of $C^{(f)}_{n(M)}$  which is determined by the multiverse (\ref{Tema46}). If
\beq
C^{(b)}_{n(M)}<C^{(f)}_{n(M)},
\label{Tema47}
\enq
then the only way for the frog to explain this difference will be by assuming the existence of some additional type of matter, which does not manifest itself in any way other than gravitationally, and is therefore deserving to be called a ``dark matter''. In other words, within the framework of our model, a completely new and unexpected approach to the explanation of the mystery of dark matter arises.

The next interesting observation concerns the vacuum energy. The values of $\rho^{(b)}_{n(\Lambda)}$ can be of different signs, therefore, certain compensation is not excluded when summing them. In other words, the vacuum energy density observed by a ``frog'' may well be less than the actual density of vacuum energy in a given universe. Of course, it is hardly possible to explain in this way the colossal difference in the prediction of quantum field theory and what we observe (the difference reaches 50 orders of magnitude). On the other hand, as is known, an anthropic restriction on the value of $\rho^{(b)}_{k(\Lambda)}$ gives the upper limit two orders of magnitude greater than the observed density of the vacuum energy.

There is a catch here. The fact is that in the presence of many biophilic parameters, the anthropic principle predicts that the observer with high probability should find herself close to the biophilic boundary. This circumstance is rarely discussed in the literature, but it is very significant. In fact, the anthropic principle leads to a conclusion, that we (as typical observers) must live literally "on the brink of the abyss" \cite{David}. Let's consider this issue in more detail.

Suppose we have $L$ biophilic parameters, which we denote by $\alpha_j$, $j=1,...,L$.  In order to permit a carbon-based life to exist, each of these parameters should belong to the predetermined intervals $\alpha'_j<\alpha_j<\alpha''_j$. If a parameter $\alpha_j$ happens to lie outside of its interval, the existence of life and the observers becomes impossible.

Let us define renormalized biophilic parameters:
$$
\beta_j=\frac{\alpha_j-\alpha'_j}{\alpha''_j-\alpha_j}.
$$
Obviously, $0<\beta_j<1$.  Now our existence in the fixed universe is  determined by the $L$-dimensional point, with coordinates $\left\{\beta_j\right\}$ lying inside the unit $L$-dimensional sphere with the volume $V_L=2\pi^{L/2}/(L\Gamma(L/2))$, where $\Gamma(L/2)$ is the Gamma-function. A striking property of multidimensional spheres is that almost all of its volume is concentrated near the surface of the sphere. In particular, for $L=10$ (but this is a modest assumption: the number of biophilic parameters is probably much greater), 65\%  of the total volume lies in the spherical layer at a distance of $0.1$ from the surface of the unit sphere.

Suppose that all biophilic coordinates uniformly fill the interior of the unit sphere. Let $\beta_1$ be the coordinate corresponding to the density of the vacuum energy. Since the observed value is ten to one hundred times smaller than the upper limit, it turns out that we are inside a sphere of radius $0.01-0.1$. For $L = 10$, the probability of finding itself inside such a sphere lies in the interval  $(10^{-20},\,10^{-10})$.

Of course, we can say that this is nothing more than a rough estimate; we can assume that the parameter $\beta_1$  accidentally turned so close to the center of the unit sphere, while the other nine parameters lie near the surface of this sphere. Nevertheless, the numbers of the order of $10^{-10}$  seem to be a serious problem. It is difficult to imagine any simple and plausible way to increase these probabilities by 9 orders of magnitude, while remaining within the framework of the classical paradigm.

However, it {\em can} be done within the framework of our model. Indeed, as follows from (\ref{Tema46}), the vacuum energy density from the point of view of a ``frog'' is determined by summing up the true (``bird's'') densities of the vacuum energy. These densities can be of different sign and therefore can well compensate each other by an order of magnitude or two, so that
$$
\rho^{(b)}_{n(\Lambda)}=(10-100)\times \rho^{(f)}_{n(\Lambda)}.
$$

And this leads us to the last important question that we should explore before we conclude this article: what would happen with our model if the vacuum energy becomes the {\em only} dominant part of the matter fields in the universe? For example, if the universe expands long enough, so that the radiation's and the baryon matter's contributions become negligible? And what if the universe is filled with the {\em negative} cosmological constant $\Lambda$? Would the Q-potential matter in these cases?

In order to answer these questions let us take a look a two important cosmological models: the de Sitter (dS) and anti-de Sitter (AdS) models.

\section{What would the frog see in the de Sitter and Anti-de Sitter universes?}\label{sec:AdS}

Suppose the universe is filled only with vacuum energy and the Q-potential, that is, $C^{(f)}_{n(M)}=C^{(f)}_{n(R)} = 0$ in (\ref{Tema45}). For simplicity, we will omit the indexes $(f)$ and $n$. Expression (45) takes the form
\beq
\frac{{\dot{R}}^2}{R^2}=\frac{8\pi G}{3}\left(\rho_{\Lambda}-\frac{3\sigma_N^2M_N^2}{4\pi G K_N^2}\frac{1}{R^4}\right)-\frac{k c^2}{R^2},
\label{Tema51}
\enq
We assume that $k =\pm  1$. The case $k = 0$  is the simplest, and we leave it to the reader (note that  $k = 0$  means $K_N = 0$, so all formulas must be redefined, for example, $K_N\to 0$ results in the divergence in the second term in parentheses in (\ref{Tema51}). Further, we separately consider the case of dS ($\rho_{\Lambda}>0$)  and AdS ($\rho_{\Lambda}<0$).
\subsection{The de Sitter universe}

Let $\rho_{\Lambda}>0$, then the solution of the equation (\ref{Tema51}) has the form
\beq
R_{dS}(t)=\sqrt{\frac{\sqrt{c^4+\nu^2}\cosh\left(2\sqrt{\Lambda} t\right)+c^2 k}{2\Lambda}},
\label{Tema52}
\enq
where
\beq
\Lambda=\frac{8\pi G\rho_{\Lambda}}{3},\qquad \nu=\frac{2\sqrt{2\Lambda} M_N\sigma_N}{\left|K_N\right|}.
\label{Tema53}
\enq

This solution does not vanish either for $k = 1$  or for $k = -1$. The Hubble parameter:
\beq
H=\frac{\sqrt{\Lambda\left(c^4+\nu^2\right)}\sinh\left(2\sqrt{\Lambda} t\right)}{\sqrt{c^4+\nu^2}\cosh\left(2\sqrt{\Lambda} t\right)+c^2 k},
\label{Tema54}
\enq
that is, we have a model with a rebound at $t=0$. Calculating the parameter of the equation of state gives us
\beq
w+1=-\frac{4}{3}\frac{\nu^2}{\left(y+kc^2\right)^2-\nu^2},
\label{Tema55}
\enq
where  $y=\sqrt{c^4+\nu^2}\cosh\left(2\sqrt{\Lambda} t\right)$. Interestingly, for $k=+1$ we get a phantom dynamics provided that
\beq
\cosh\left(2\sqrt{\Lambda} t\right)>\frac{\nu-c^2}{\sqrt{c^4+\nu^2}}.
\label{Tema56}
\enq
Note that
\beq
\nu-c^2=c^2\left(\frac{\sqrt{2\Lambda} M_NZ_N t_{PL}}{\left|K_N\right|}-1\right).
\label{Tema57}
\enq
If the right-hand side of (\ref{Tema57}) is negative, then inequality (\ref{Tema56}) holds over the entire time interval: $t\in (-\infty,+\infty)$. In other words, the universe always exists in a phantom mode, although (except at $t\sim 0$) the parameter $w$ is practically indistinguishable from the value of negative one: $w=-1-\epsilon$, $\epsilon\ll 1$. Surprisingly, a similar situation occurs with $\nu>c^2$, because for any real and positive $\nu$:
$$
\left(\nu-c^2\right)^2>c^4+\nu^2.
$$
Thus, for $k = +1$, the scale factor (\ref{Tema52}) describes the phantom regime.

In the case of an open universe ($k=-1$), the situation is slightly different. If $\nu>c^2$ then the  phantom regime  is realized if
\beq
\cosh\left(2\sqrt{\Lambda} t\right)>\frac{\nu+c^2}{\sqrt{c^4+\nu^2}}>1.
\label{Tema58}
\enq
The dynamics of the process is as follows: during the compression phase, the universe almost always remains in a phantom state. Before it reaches the minimum value of $a$ (the rebound point), there is a crossing of the boundary of the phantom zone. Then a rebound occurs and at the expansion stage there is a second crossing of the phantom boundary. After that, the universe is constantly expanding in phantom mode, but the dynamics very quickly converge to a standard dS mode,  with  $w=-1-\epsilon$, $\epsilon\ll 1$.

If $\nu < c^2$  then phantoms take place in regions: $y<y_{-}=c^2-\nu$  and   $y>y_{+}=c^2+\nu$.  Since the smallest value of $y$ is $y_{min}=y(t=0)=\sqrt{c^4+nu^2}$ and for any positive $\nu$, $y_{min}>y_{-}$, one conclude that  the phantoms will exist only in the region $y>y_{+}$ and  this condition is equivalent to (\ref{Tema58}). Thus, in the open universe, regardless of the magnitude of the $\nu>0$, there is always a double crossing of the boundary of the phantom zone.

\subsection{The Anti-de Sitter universe}

We now turn out attention to the case of a negative vacuum energy; $\rho_{\Lambda}<0$. Let $\Lambda=8\pi G\left|\rho_{\Lambda}\right|/3$, and define $\nu$ as in (\ref{Tema53}). Obviously, the equation (\ref{Tema51}) makes sense only in the open universe. Integrating it one has
\beq
R_{AdS}(t)=\sqrt{\frac{c^2+\sqrt{c^4-\nu^2}\sin\left(2\sqrt{\Lambda} t\right)}{2\Lambda}},
\label{Tema510}
\enq
from which we immediately obtain an interesting restriction
\beq
\left|\rho_{\Lambda}\right|<\frac{3  K_N^2}{64\pi G M_N^2 Z_N^2 t^2_{PL}}.
\label{Tema511}
\enq
We emphasize that in the case of a positive vacuum energy density, we do not obtain any restrictions on the value of this density.

The Hubble parameter
\beq
H=\frac{\sqrt{\Lambda\left(c^4-\nu^2\right)}\cos\left(2\sqrt{\Lambda} t\right)}{c^2+\sqrt{c^4-\nu^2}\sin\left(2\sqrt{\Lambda} t\right)}.
\label{Tema512}
\enq
We note that $R_{AdS}(t)$  does not vanish anywhere, but oscillates between
\beq
R^{min/max}_{AdS}=\sqrt{\frac{c^2\mp \sqrt{c^4-\nu^2}}{2\Lambda}}.
\label{Tema513}
\enq
It  is easy to see that
\beq
w+1=\frac{4\nu^2}{3\left(\nu^2+4\Lambda^2 R^4_{AdS}\right)}>0,
\label{Tema515}
\enq
therefore we have no phantoms in AdS model at all. Nevertheless, the universe is experiencing accelerated expansion at certain time intervals.

Indeed,
\beq
\frac{{\ddot{R}}_{AdS}(t)}{R_{AdS}(t)}=\frac{\nu^2-4\Lambda^2R^4_{AdS}(t)}{4\Lambda R^4_{AdS}(t)}.
\label{Tema516}
\enq

If we make a substitution $c^2=A\cosh\phi$, $\nu=A\sinh\phi$  then condition (\ref{Tema511}) will be fulfilled automatically. Now lets substitute $R^{min}_{AdS}$ (see (\ref{Tema513})  instead of $R_{AdS}(t)$  in the numerator of the right-hand side of equation (\ref{Tema516}). As a result  this numerator will be equal to
$$
2A^2\left(\cosh\phi-1\right)>0,
$$
but if we substitute $R^{max}_{AdS}$  then this numerator will be
$$
-2A^2\left(\cosh\phi+1\right)<0.
$$

Thus we arrive at the following picture: the first part of the expansion of the universe (from $R^{min}_{AdS}$) comes with acceleration. The second part slows down to a full stop (at $R_{AdS}=R^{max}_{AdS}$), after which compression begins with the opposite sign change in acceleration.

This is extremely interesting as it is quite an unusual behaviour for the models with a negative vacuum energy density. Of course, the cause of acceleration (at the stage of expansion) is nothing else but our Q-potential. This observation opens up an intriguing opportunity. Until now, the the accelerated expansion of the universe was considered an unambiguous indication that the cosmological constant is positive. As we see, the presence of a Q-potential can dramatically change this belief. Given that the negative cosmological constant agrees much better with string theory, we can express a timid hope that the introduction of a Q-potential into cosmology can prove very useful for string physics.

\section{The Discussion} \label{sec:discussion}

In this article we have aimed to demonstrate that the method of MIW quantization can be effectively generalized for the Friedmann cosmologies. Let us now briefly sum up the most important obtained results so far.

\begin{enumerate}
\item The inclusion of the quantum potential (Q-potential) into the Friedmann framework leads to a natural emergence of phantom dynamics, albeit without the pathological effects like the Big Rip singularities. The phantom phase can be either limited in time or be infinite in length, the particulars strongly depending  on the types of the matter fields that fills the universe.

\item In some cases the Q-potential generates the $w$-singularities.

\item The application of MIW in cosmology allows for two distinct descriptions: from the point of view of the internal observer (the ``frog'') and the overall view over the entire ensemble of universes (the ``bird''). Interestingly, in order to satisfy these two descriptions together (via the master-factor), the ``frog'' might require a concept of ``dark matter'' just to explain the cosmological dynamics it observes.

 \item From the point of view of a ``frog'', living in the universe with a dominant positive cosmological constant, the parameter of state should look like $w=-1-\epsilon$, with positive $\epsilon \ll 1$. Curiously, this seems to be consistent with the most recent estimates on $w$.

 \item For a ``frog'', who lives in a universe with a dominant negative cosmological constant, the Q-potential leads to a very peculiar and unexpected effect: the universe ends up being cyclically oscillating in time. This is made possible by two facts: first, the collapse of the universe ends up not in singularity, but in rebound from a small value of scale factor. Second, for a finite time after the rebound a new expansion begins, during which we have $\ddot H >0$: a most unusual behaviour for the models with the negative cosmological constant.

\end{enumerate}

What lies ahead now? Before we finish, we would like to discuss the future prospects of our method and the immediate possible developments that it allows for.

\begin{enumerate}
\item One of the first and most immediate steps would be to test the models with Q-potential on existing astronomical observational data. The preliminary tests we have conducted so far have shown that the $\Lambda C D M$-model equipped with a Q-potential appears to work quite satisfactorily.

\item Another very interesting area of investigation is the physics of the wormholes in the universe with Q-potential. It is a known fact that in the presence of the phantom fields the stability and even the growth of the wormholes changes rather dramatically, specifically due to accretion of the phantom energy onto the wormhole \cite{Pedro_2004}. While the particular dynamics is still open for a debate (see, for example, \cite{FI} and \cite{Pedro_2006}), it is indisputable that the phantoms must play a tremendous role in the wormhole's evolution. Our preliminary estimates show that in the dS universe filled with Q-potential the radius of the wormholes would indeed grow, even though the phantom phase is rather limited in time.

\item In a series of articles (see, for example, \cite{Tipler}) it has been claimed that the proper ``quantization of cosmology'' has to be performed in a conformal time. It can be shown that the MIW formalism can also be easily adapted in this case. However, it leads to the models that are rather different from those discussed in this article. In particular, in the framework of the master-factor method the Q-potential appears to be decreasing much faster with the scale factor. This means, that the corresponding models have to be studied  and tested separately, being essentially the alternatives to the models we have described so far. Once again, according to the preliminary analysis, done for the universe with a dominant positive cosmological constant, these models might stick even closer to the observational data, and also allow for the ``growth of the wormholes'' phenomena. At the same time, it appears that all models with negative cosmological constant predict a rebound from a collapse, even if the universe is also filled with radiation.

\item Finally, a very interesting question would be an application of our method to the models more general then simple Friedmann cosmologies. It is known that the final singularity that arises in even slightly non-isotropic collapsing universe is radically different from the initial singularity, as it is characterized by a regime of oscillations, with rapid alteration of Kasner epochs. In order to understand the role of Q-potential in this circumstances (and find a definitive answer to a very important question of whether it might lead to a {\em rebound} from the singularity) it would be imperative to generalize the MIW method to the anisotropic models. It is too early to claim anything with certainty, but the preliminary calculations demonstrate that while the complete rebound might not occur, a very new types of singularity arise, in which the universe collapses along two out of three of ellipsoid's axes, while infinitely expanding, Big Rip-style, along the third one.
\end{enumerate}

These are four main areas that are currently being investigated by our group. The preliminary results that we have mentioned above are being prepared as a subject of the next article on this incredibly fascinating subject.

\begin{acknowledgments}
This article is dedicated to the memory of our dear friend and colleague, Pedro Gonz\'alez-D\'iaz.

The work was supported and funded by the project 1.4539.2017/8.9 (MES, Russia).
\end{acknowledgments}

\end{document}